\documentclass[preprint,aps,amstex]{revtex4}

\begin{document}

\title{Quantum Zeno and anti-Zeno Effects:  An Exact Model}

\author{T. C. Dorlas and R. F. O'Connell$^{\dagger}$}
\affiliation{School of Theoretical Physics, Dublin Institute for Advanced Studies, 10 Burlington
Road, Dublin 4, Ireland}
\date{\today }

\begin{abstract}
Recent studies suggest that both the quantum Zeno (increase of the natural lifetime of an
unstable quantum state by repeated measurements) and anti-Zeno (decrease of the natural lifetime)
effects can be made manifest in the same system by simply changing the dissipative decay rate
associated with the environment.  We present an {\underline{exact}} calculation confirming this
expectation.
\\
\\
\\
\\
\\
\\
\\
\\
\\
\\
\\
\\
\\
\\
\\

\noindent PACS number(s):  03.65.Xp, 03.65.Yz, 05.30.-d, 05.40.Jc 
\end{abstract}


\maketitle

The quantum Zeno effect (QZE) predicts that the lifetime of an excited state increases by
repeated measurements.  It has been a subject of interest for many years \cite{beskow,misra} and
recent reviews appear in \cite{home} and \cite{facchi}.  More recently, it has been pointed out
that a decrease in lifetime, referred to as the Inverse or anti-Zeno effect (AZE) can also
occur
\cite{pascazio,kofman,luis}.  Whereas there are claims that the QZE has been observed, to our
knowledge there has been no experimental verification yet of the AZE.  On the other hand, the
detailed calculations of Pascazio and Facchi \cite{pascazio} and Kofman and Kurizki \cite{kofman}
lend strong credence to the possible existence of the AZE.  Since the calculations of
\cite{pascazio} and \cite{kofman} by their nature required various assumptions (such as the
Weisskopf-Wigner approximation), we consider it desirable to present an {\underline{exact}}
calculation which should also delineate in a clear-cut manner the nature and magnitude of the
external environment that is necessary to achieve the transition from QZE to AZE.

The system we analyze is the decay of a free particle that is placed initially in a Gaussian
state:

\begin{equation}
\psi (x,0)=\left(2\pi\sigma^{2}\right)^{-1/4}\exp\left\{-\frac{x^{2}}{4\sigma^{2}}\right\},
\label{qzaz1}
\end{equation}
where $\sigma^{2}$ is the variance.  The particle is regarded as part of a larger system of a
particle coupled to a reservoir and the complete system is initially in equilibrium at
temperature $T$.  This was the scenario considered by Ford et al. \cite{ford1} who used
distribution functions defined in accordance with the quantum theory of measurement to obtain
exact results for the spreading of the wave packet and for the probability at time $t$ given by

\begin{equation}
P(x,t)=\frac{1}{\sqrt{2\pi w^{2}(t)}}~\exp\left\{-\frac{x^{2}}{2w^{2}(t)}\right\}. \label{qzaz2}
\end{equation}
Here

\begin{eqnarray}
w^{2}(t) &=& \sigma^{2}-\frac{\left[x(0),x(t)\right]^{2}}{4\sigma^{2}}+s(t) \nonumber \\
&\equiv& \sigma^{2}+\sigma^{2}_{q}+s(t), \label{qzaz3}
\end{eqnarray}
where $\sigma^{2}$ is the initial variance, $\left[x(0),x(t)\right]$ is the commutator,

\begin{equation}
s(t)=\left\langle\left\{x(t)-x(0)\right\}^{2}\right\rangle , \label{qzaz4}
\end{equation}
is the mean square displacement and $\sigma^{2}_{q}$ is the contribution to the spreading due to
temperature-independent quantum effects.  A measure of the decay rate $R(t)$ is simply given by the ratio
of the probabilities at times $t$ and $0$.  However, this ratio is clearly dependent on $x$ so, from
henceforth, we take $x=0$ (corresponding to the maximum of the wave-packet) and write

\begin{eqnarray}
R(t) &=& \frac{P(0,t)}{P(0,0)} \nonumber \\
&=& \left\{\frac{\sigma^{2}}{w^{2}(t)}\right\}^{1/2}. \label{qzaz5}
\end{eqnarray}
Hence, our calculation reduces to an evaluation of the width of the wave-packet at time $t$.

The quantities appearing in (\ref{qzaz3}) and (\ref{qzaz4}) are evaluated by use of the quantum
Langevin equation \cite{ford2}, which is a Heisenberg equation of motion for $x(t)$, the
dynamical variable corresponding to the coordinate of a Brownian particle interacting with a
linear passive heat bath.  For the case of a free particle, this equation for the stationary
process has the well known form,

\begin{equation}
m\ddot{x}+\int^{t}_{-\infty}dt^{\prime}\mu (t-t^{\prime})\dot{x}(t^{\prime})=F(t), \label{qzaz6}
\end{equation}
where $\mu (t)$ is the memory function and $F(t)$ is a fluctuating operator force with mean
zero.  The solution of the quantum Langevin equation (\ref{qzaz6}) can be written

\begin{equation}
x(t)=\int^{t}_{-\infty}dt^{\prime}G(t-t^{\prime})F(t^{\prime}), \label{qzaz7}
\end{equation}
where $G(t)$, the Green function, can in turn be written

\begin{equation}
G(t)=\frac{1}{2\pi}\int^{\infty}_{-\infty}d\omega\alpha (\omega+i0^{+})e^{-i\omega t},
\label{qzaz8}
\end{equation}
in which $\alpha (z)$ (the Fourier transform of the Green function) is the response function. 
For the free particle the response function has the general form

\begin{equation}
\alpha (z)=\frac{1}{-mz^{2}-iz\tilde{\mu}(z)}, \label{qzaz9}
\end{equation}
in which $\tilde{\mu}(z)$ is the Fourier transform of the memory function,

\begin{equation}
\tilde{\mu}(z)=\int^{\infty}_{0}dt\mu (t)e^{izt}, ~~~~{\textnormal{Im}}\{z\}>0. \label{qzaz10}
\end{equation}

Using these results, we find that \cite{ford2,ford3} the mean square displacement is given by the
formula

\begin{equation}
s(t)=\frac{2\hbar}{\pi}\int^{\infty}_{0}d\omega{\textnormal{Im}}\left\{\alpha \left(\omega +i0^{+}
\right)\right\}\coth\frac{\hbar\omega}{2kT}(1-\cos\omega t), \label{qzaz11}
\end{equation}
while the commutator, which is temperature independent, is given by the formula

\begin{equation}
[x(0),x(t)]=\frac{2i\hbar}{\pi}\int^{\infty}_{0}d\omega{\textnormal{Im}}\left\{\alpha
\left(\omega +i0^{+}\right)\right\}\sin\omega t. \label{qzaz12}
\end{equation}
These expressions are valid for arbitrary temperature and arbitrary dissipation.  (Indeed, with
the appropriate expression for the response function, they are valid in the presence of an
external oscillator potential.)  Here, we confine our attention to the case of zero temperature
and Ohmic dissipation, where $\tilde{\mu}(z)=m\gamma$.  It then follows that

\begin{equation}
s(t)=\frac{2\hbar\gamma}{\pi m}t^{2}{\textnormal{I}}(\gamma t), \label{qzaz13}
\end{equation}
where

\begin{equation}
{\textnormal{I}}(\gamma t)=\int^{\infty}_{0}dy\frac{(1-\cos y)}{y\left[y^{2}+(\gamma
t)^{2}\right]}. \label{qzaz14}
\end{equation}
In addition, the commutator is given by

\begin{equation}
\left[x(0),x(t)\right]=i\hbar G(t)=\frac{i\hbar}{m\gamma}\left(1-e^{-\gamma t}\right),
\label{qzaz15}
\end{equation}
so that

\begin{equation}
\sigma^{2}_{q}=\frac{\hbar^{2}}{m^{2}\gamma^{2}}~~\frac{\left(1-e^{-\gamma
t}\right)^{2}}{4\sigma^{2}}. \label{qzaz16}
\end{equation}
Hence, we now have all the tools at our disposal in order to carry out an exact calculation  of
$P(x,t)$ and hence the rate $R(t,\gamma )$, where we have added the argument $\gamma$ to $R$ in
order to emphasize the fact that this dependence will be the crucial element in our calculation. 
Combining the various results given above, we may write explicitly

\begin{equation}
R(t,\gamma
)=\left\{\left[\sigma^{2}+\left(\frac{\hbar^{2}}{4m^{2}\sigma^{2}}\right)t^{2}\left\{\frac
{\left(1-e^{-\gamma t}\right)}{\gamma t}\right\}^{2}+\frac{2\hbar t}{\pi m}(\gamma
t){\textnormal{I}}(\gamma t)\right]/\sigma^{2}\right\}^{-1/2}. \label{qzaz17}
\end{equation}

Our goal will be to calculate $\left\{R(\tau,\gamma)\right\}^{n}$, the rate corresponding to $n$
measurements on the system, where $\tau =t/n$, and then compare it to $R(t,\gamma)$.  As we
shall see, the result depends crucially on the value of
$\gamma t$.

The only quantity left requiring explicit evaluation is ${\textnormal{I}}(\gamma t)$ given by
(\ref{qzaz14}).  In order to obtain more physical insight into the nature of the results
obtained, we will first evaluate $R$ analytically for both small and large values of $\gamma t$,
which we will demonstrate correspond to the QZE and AZE, respectively.  However, in order to
determine for what value of $\gamma t$ the transition between the two regimes occur, it will be necessary
to carry out a numerical evaluation of
${\textnormal{I}}(\gamma)$.  First, we turn to the analytic calculation.

\noindent (a) {\underline{$\gamma t<<1$}} \\
\noindent Then, from (\ref{qzaz16}),

\begin{equation}
\sigma^{2}_{q}\approx\frac{\hbar^{2}}{4m^{2}\sigma^{2}}t^{2}, \label{qzaz18}
\end{equation}
which corresponds to the usual dynamical wave packet spreading in the absence of a dissipative
environment.  In addition, (\ref{qzaz13}) reduces to \cite{ford4}

\begin{equation}
s(t)=\frac{\hbar\gamma}{\pi m}t^{2}\left\{-\log(\gamma t)+\frac{3}{2}-\gamma_{E}\right\},
\label{qzaz19}
\end{equation}
where $\gamma_{E}=0.577$ is Euler's constant.

Using these results in (\ref{qzaz3}) leads to

\begin{equation}
w^{2}(t)=\sigma^{2}+\langle v^{2}\rangle t^{2}, \label{qzaz20}
\end{equation}
where

\begin{equation}
\langle v^{2}\rangle=\frac{\hbar^{2}}{4m^{2}\sigma^{2}}+\frac{\hbar\gamma}{\pi
m}\left\{-\log(\gamma t)+0.92\right\}. \label{qzaz21}
\end{equation}
Since for most reasonable scenarios $\langle v^{2}\rangle t^{2}<<\sigma^{2}$, we may expand $w^{2}(t)$ in
a power series in $t^{2}$ to get

\begin{eqnarray}
R(t) &=& \left\{\frac{\sigma^{2}}{w^{2}(t)}\right\}^{1/2}\approx
1-\frac{1}{2\sigma^{2}}\left\{\sigma^{2}_{q}+s(t)\right\}
\nonumber
\\ &\approx& \left(1-\frac{\langle v^{2}\rangle t^{2}}{2\sigma^{2}}+ - - -\right).
\label{qzaz22}
\end{eqnarray}
Apart from the weak $\log (\gamma t)$ dependence of $\langle v^{2}\rangle$, as manifest in
(\ref{qzaz21}), we note the ubiquitous short-time $t^{2}$ behaviour of $R(t)$, which is
characteristic of the QZE.  To remind the reader, we consider $n$ instantaneous measurements at
intervals $\tau =t/n$.  Thus, for large $n$

\begin{eqnarray}
\left[R(\tau )\right]^{n} &=& \left(1-\frac{\langle v^{2}\rangle t^{2}}{n}~\frac{1}{n}\right)^{n}
\nonumber \\
&=& \exp\left(-\langle v^{2}\rangle t\tau\right) \nonumber \\
&\rightarrow&1~~{\textnormal{as}}~~\tau\rightarrow 0. \label{qzaz23}
\end{eqnarray}
In other words, for small $\gamma t$, the decay rate becomes frozen as $n\rightarrow\infty
~(\tau\rightarrow 0)$, which is the extreme manifestation of the QZE.  Moreover, for
$\gamma\rightarrow 0$, we see from (\ref{qzaz21}) that the second term on the right-side goes
to zero.  Thus, (\ref{qzaz23}) still holds but now $\langle v^{2}\rangle$ is completely
independent of both $\gamma$ and $t$.  The conclusion is that in the absence of a heat bath,
the QZE always holds.

\noindent (b)  {\underline{$\gamma t>>1$}} \\
\noindent Then, from (\ref{qzaz16}),

\begin{equation}
\sigma^{2}_{q}\approx\frac{\hbar^{2}}{4m^{2}\sigma^{2}}~\frac{1}{\gamma^{2}}, \label{qzaz24}
\end{equation}
and (\ref{qzaz13}) reduces to \cite{ford4}

\begin{equation}
s(t)=\frac{2\hbar}{\pi m\gamma}\left\{\log (\gamma t)+\gamma_{E}\right\}. \label{qzaz25}
\end{equation}
Hence

\begin{equation}
w^{2}(t)=\sigma^{2}+\frac{\hbar^{2}}{4m^{2}\sigma^{2}}~\frac{1}{\gamma^{2}}+\frac{2\hbar}{\pi
m\gamma}\left[\log (\gamma t)+\gamma_{E}\right]. \label{qzaz26}
\end{equation}
Since for most reasonable scenarios, $\sigma^{2}>>s(t)>>\sigma^{2}_{q}$ for any $t$, we may
write

\begin{eqnarray}
R(t) &\approx& 1-\frac{s(t)}{2\sigma^{2}} \nonumber \\
&=& 1-\frac{\hbar}{\pi m\sigma^{2}\gamma}\left[\log (\gamma t)+\gamma_{E}\right].
\label{qzaz27}
\end{eqnarray}
Because of the weak $\log$ dependence on $t$, it is clear that, for fixed $\gamma$,
$R\left(\frac{t}{n}\right)$ is just slightly larger than $R(t)$, keeping in mind that
one must be careful to ensure that $\gamma\tau >>1$ based on our initial assumption.  Thus,
it is readily apparent that

\begin{eqnarray}
\left[R(\tau )\right]^{n} &<& \left\{1-\frac{\hbar}{\pi
m\sigma^{2}\gamma}\left[\log\left(\frac{\gamma t}{n}\right)+\gamma_{E}\right]\right\}^{n}
\nonumber \\
&<& R(t) . \label{qzaz28}
\end{eqnarray}
Thus, we see
that
$\left[R(\tau )\right]^{n}$ is always less than $R(t)$ and decreases with increasing $n$
(decreasing
$\tau$), which is the extreme manifestation of the AZE.  Moreover, for $n$ sufficiently
small, we may write

\begin{equation}
\left[R(\tau )\right]^{n} \approx \exp\left\{-\frac{\hbar n}{\pi
m\sigma^{2}\gamma}\left[\log\left(\frac{\gamma t}{n}\right)+\gamma_{E}\right]\right\},
\label{qzaz29}
\end{equation}
the restriction on $n$ being to ensure that the magnitude of the argument of the exponential
is $<<1$.

The conclusion is that the QZE is characterized by small $\gamma t$ values whereas the AZE is
characterized by large $\gamma t$ values.  As a check on the analytic results give for
$I(\gamma t)$ in (\ref{qzaz14}), we carried out a numerical evaluation of the integral and
obtained excellent agreement.

In order to obtain the value of $\gamma t$ for which
the transition occurs, as well as delineating more accurately the analytic results obtained
above, we now turn to a numerical evaluation of $R(t,\gamma )$ given in (\ref{qzaz17}).   
Thus, in Figs 1 and 2, we plot
$\left\{R\left(\frac{t}{20}\right)\right\}^{20}$, corresponding to 20 measurements, and compare
it to $R(t)$, for $t$ values ranging from 0-5 and 0-400, respectively, and taking $\gamma =0.1$. 
We note that $\left\{R\left(\frac{t}{20}\right)\right\}^{20}$ is initially larger than $R(t)$,
corresponding to the QZE but it becomes smaller (corresponding to the AZE) for $\gamma t$ values
larger than the transition $\gamma t$ value of ? ? .

In conclusion, we have presented an {\underline{exact}} calculation of the decay rate of a free
particle that is placed initially in a Gaussian state and which is coupled to a reservoir so that
the complete system in initially in equilibrium at zero temperature.  The results obtained
demonstrate that repeated measurements made on the system lead to a QZE effect scenario for small
$\gamma t$ values while evolving into an AZE effect scenario for large $\gamma t$ values,
confirming similar results obtained in \cite{pascazio} and \cite{kofman}.

\acknowledgements

RFO'C would like to thank the School of Theoretical Physics, Dublin Institute for
Advanced Studies, for their hospitality while this work was being carried out.

\end{document}